\documentclass[aps,superscriptaddress,nofootinbib,preprintnumbers]{revtex4}
\usepackage{amsmath}
\usepackage{graphicx,tabularx}
 
\newcommand{\sla}[1]{/\!\!\!#1}

\newcommand{\p}{\mathrm{p}}

\newcommand{\go}   {\tilde{g}}

\newcommand{\sbx}[1]{\tilde{b}_{{#1}}}
\newcommand{\sq}[1]{\tilde{q}_{{#1}}}

\newcommand{\se}[1]{\tilde{\ell}_{{#1}}}

\newcommand{\smuon}[1]{\tilde{\mu}_{{#1}}}
\newcommand{\nn}[1]{\tilde{\chi}^0_{{#1}}}

\providecommand{\SP}{\scriptscriptstyle}

\newcommand{\mst}[1]{m_{\tilde{t}_{\SP {#1}}}}
\newcommand{\msb}[1]{m_{\tilde{b}_{\SP {#1}}}}

\newcommand{\mnn}[1]{m_{\tilde{\chi}^{\SP 0}_{\SP {#1}}}}

\def\ie{{\it i.e.\;}}
\def\eg{{\it e.g.\;}}

\newcommand{\gev}{~{\ensuremath\rm GeV}}
\newcommand{\tev}{~{\ensuremath\rm TeV}}

\newcommand{\ifb}{~{\ensuremath\rm fb^{-1}}}
 
\begin{document}
 
\date{\bf 02/04/07}
 
\title{Unravelling the sbottom spin at the CERN LHC}


\author{Alexandre Alves}
\email{aalves@fma.if.usp.br}
\affiliation{Instituto de F\'{\i}sica, Universidade de S\~{a}o Paulo,
             S\~{a}o Paulo, Brazil}
\author{Oscar \'Eboli}
\email{eboli@fma.if.usp.br}
\affiliation{Instituto de F\'{\i}sica, Universidade de S\~{a}o Paulo,
             S\~{a}o Paulo, Brazil}

\begin{abstract}

\bigskip

Establishing that a signal of new physics is undoubtly supersymmetric requires
not only the discovery of the supersymmetric partners but also probing their
spins and couplings. We show that the sbottom spin can be probed at the CERN
Large Hadron Collider using only angular correlations in the reaction
$pp \to \sbx{} \sbx{}^* \to b \bar{b} \sla{p}_T$, which allow us to
distinguish a universal extra dimensional interpretation with a fermionic
heavy bottom quark from supersymmetry with a bosonic bottom squark. We
demonstrate that this channel provides a clear indication of the sbottom spin
provided the sbottom production rate and branching ratio into $b \nn{1}$ are
sufficiently large to have a clear signal above Standard Model backgrounds.

\end{abstract}
 
\maketitle
 
\section{Introduction}
\label{sec:intro}

Supersymmetric models are promising candidates for physics beyond the Standard
Model (SM), despite the present lack of direct experimental evidence on
supersymmetry (SUSY). The CERN Large Hadron Collider (LHC) has a large reach
for the discovery of SUSY~\cite{susy} that largely relies on the production
and decay of strongly interacting new particles {\em i.e.}  squarks and
gluinos~\cite{deq,prospino}.  Establishing that a signal of new physics at the
LHC is indeed supersymmetric requires not only the discovery of the new
supersymmetric partners but also probing their interactions and
spins~\cite{barr1,barr2,smillie,aet}.  \medskip

Previously, the squark spin has been studied through the long decay chain $
\sq{} \to \nn{2} \to \se{} \to \nn{1}$ which is also used to measure its
mass~\cite{barr1,smillie} or in conjunction with the gluino spin analysis in
the decay chain $ \go \to \sbx{} \to \nn{2} \to \se{} \to \nn{1}$~\cite{aet}.
In this work we probe the potential of the CERN Large Hadron Collider (LHC)
for unravelling the bottom squark spin using angular correlations in the short
decay chain $pp \to \sbx{} \sbx{}^* \to b \bar{b} \sla{p}_T$, analogously to
what has been done for the analysis of the slepton spin~\cite{barr1,barr2}.  A
nice feature of this reaction is that only sbottoms and the lightest SUSY
particle (LSP) takes place in it, providing a further check of sbottom spin
obtained in the long decay chain studies.  Unfortunately, this analysis can
not be extended straightforwardly to light flavor squarks due to large QCD
backgrounds.
\medskip 

To determine the spin nature of the sbottoms at the LHC we compare the SUSY
sbottom production and decay chain with another scenario where the new
intermediate states have the same spin as the SM particles and leads to the
same final state $b \bar{b} \sla{p}_T$. Such a model are Universal Extra
Dimensions (UED)~\cite{ued} where each SM particle has a heavy Kaluza--Klein
(KK) partner which can mimic the SUSY cascade decay, provided we employ the
mass spectra extracted from the decay kinematics match~\cite{early}.  Here we
are not focusing on UED searches but we use UED only for comparison with the
SUSY predictions.
\medskip 

There are many observables which we can use to discriminate `typical' UED and
SUSY models, like the production rate or the mass spectrum.  Nevertheless, at
the LHC we measure only production cross sections times branching ratios, and
the UED as well as the SUSY mass spectra are unlikely to be what we currently
consider `typical'. On the other hand, spin information is generally extracted
from angular correlations.  Therefore, we base our analysis exclusively on
distributions of the outgoing SM $b$ quarks as predicted by UED and by SUSY.
We demonstrate that this final state $b \bar{b} \sla{p}_T$ provides a clear
indication of the sbottom spin provided the sbottom production rate and
branching ratio into $b \nn{1}$ are sufficiently large to have a clear signal
above SM backgrounds.

\bigskip

\section{UED interactions and parameters}
\label{sec:ued}

We assumed one extra dimension with size $R$, where all SM fields
propagate~\cite{ued,early}, leading to a tower of discrete KK excitations
for each of the SM fields ($n=0$). In this scenario, the 5-dimensional wave
functions for an SU(2)--doublet fermion are
\begin{equation}
\psi_d = \frac{1}{\sqrt{2\pi R}} \psi^{(0)}_{dL} 
       + \frac{1}{\sqrt{\pi R}} \sum_{n=1}^{\infty}
         \left( \psi^{(n)}_{dL} \cos\frac{ny}{R}
               +\psi^{(n)}_{dR} \sin\frac{ny}{R}
         \right) \; .
\end{equation}
On the other hand, the roles of the left and right handed $n$-th KK
excitations are reversed for SU(2) singlets. Just like in the MSSM, the
spinors of the singlet ($q$) and doublet ($Q$) KK--fermion mass eigenstates
can be expressed in terms of the SU(2) doublet and singlet fields $\psi_{d,s}$
\begin{alignat}{5}
Q^{(n)} &=  \; \cos \alpha^{(n)} \psi_d^{(n)} 
             &+\;& \sin\alpha^{(n)} \psi_s^{(n)} \; , \notag \\
q^{(n)} &=  \; \sin\alpha^{(n)}\gamma^5 \psi_d^{(n)} 
             &-&   \cos\alpha^{(n)}  \gamma^5\psi_s^{(n)} \; .
\label{eq:eigen}
\end{alignat}
In general, the mixing angle $\alpha^{(n)}$ is suppressed by the SM fermion
mass over the KK--excitation mass plus one--loop corrections except for the
top quark due to its large mass.
\begin{equation}
\tan 2\alpha^{(n)}= \frac{m_f}{\frac{n}{R} +
 \frac{1}{2} (\delta m^{(n)}_Q+\delta m^{(n)}_q)} 
\label{eq:mixing1}
\end{equation}
The non--degenerate KK--mass terms $\delta m^{(n)}$ contain tree level and
loop contributions to the KK masses, including possibly large contributions
from non--universal boundary conditions.
\medskip

The neutral KK gauge fields will play the role of neutralinos in the
alternative description of the sbottom production and decay. Just as in the
SM, there is a KK--weak mixing angle which for each $n$ rotates the
interaction eigenstates into mass eigenstates
\begin{alignat}{5}
\gamma_\mu^{(n)} &=  & \cos \theta_{w}^{(n)} \, B_\mu^{(n)} 
                 &+\;& \sin \theta_{w}^{(n)} \, W_{3,\mu}^{(n)}
\; ,\notag \\
Z_\mu^{(n)}      &= -& \sin \theta_{w}^{(n)} \, B_\mu^{(n)}
                 &+&   \cos \theta_{w}^{(n)} \, W_{3,\mu}^{(n)} \; .
\label{eq:mixing2}
\end{alignat}
The $n$-th KK weak mixing angle is again mass suppressed
\begin{equation}
  \tan 2\theta_w^{(n)} = \frac{v^2 \, g \, g_Y/2}{
                         ( \delta m^{(n)}_{W_3} )^2
                       - ( \delta m^{(n)}_B )^2
                       + v^2 \left( g^2-g^2_Y \right)/4} 
\label{eq:weinberg}
\end{equation} 
where $\delta m^{(n)}$ contains tree level as well as loop corrections to the
KK gauge boson masses. Generally $(\delta m^{(n)}_{W_3} )^2 - ( \delta
m^{(n)}_B )^2 \gg v^2 \left( g^2-g^2_Y \right)$~\cite{early} and the lightest
KK partner is the $B^{(1)}$, with basically no admixture from the heavy
$W_3^{(1)}$. 
\medskip 
     
We only considered the first set of KK excitations to formulate an alternative
interpretation of production and decay of sbottoms at the LHC, using the UED
decay $ b^{(1)} \to b \gamma^{(1)}$ to mimic a sbottom decay. The relevant
interactions for this decay are
\begin{equation}
{\cal L}_{\gamma_1 q_1 q} 
= ig_Y\Big[ \;  
\bar{q}^{(0)}\gamma^\mu
                          \left( Y_d\cos\alpha^{(1)} P_L
                                +Y_s\sin\alpha^{(1)} P_R
                          \right) Q^{(1)} 
          -\bar{q}^{(0)}\gamma^\mu
                          \left( Y_d\sin\alpha^{(1)} P_L
                                +Y_s\cos\alpha^{(1)} P_R
                          \right) q^{(1)}
\Big]\gamma^{(1)}_\mu \; .
\label{eq:lagrangian1}
\end{equation} 

In general, the KK partners of the SM particles do not have a mass spectrum
similar to what we expect in SUSY, however, we imposed that the first KK
excitations have the same mass as the SUSY particles. Assigning the LSP mass
to the Lightest KK Particle (LKP) and the Next-LSP mass to the Next-LKP mass
fixes the KK-weak mixing angle by means of Eq.~(\ref{eq:weinberg})
\begin{equation}
\theta_w^{(1)} =
\frac{1}{2}\arctan\left(\frac{gg_Yv^2}{2(m^2_{NLSP}-m^2_{LSP})}\right) \; .
\label{eq:mixing3}
\end{equation}

\section{Event simulation and test points}
\label{sec:test}

We considered three scenarios for the new particle spectrum. Our first test
point is the LHC point 5 (S5) \cite{S5} that exhibits rather heavy sbottoms
with sizeable decays into $b \nn{1}$; see Table~I. Our second reference point
is SPS1a \cite{sps} where the sbottom masses are close to ones in the first
scenario, however the decays $\sbx{1,2} \to b \nn{1}$ are suppressed. The
third test point exhibits a somewhat light squark spectrum which could be
eventually produced at a 1 TeV International $e^+ e^-$ Linear Collider; we
denote this point by L1.  The masses for the L1 parameter point are $\msb{1} =
280 \gev$, $\msb{2} = 354 \gev$, $\mst{1} = 339 \gev$, $\mst{2} = 371 \gev$,
and $\mnn{1} = 97.8 \gev$. For the SPS1a and S5 parameter choices the lighter
of the two sbottoms is almost entirely a left state $\sbx{1}\sim \sbx{L}$
while for L1 it is approximately a maximal mixture of left and right states
$\sbx{1}\sim \sqrt{2}/2\sbx{L}+\sqrt{2}/2\sbx{R}$.  The salient features of
these reference points is summarized in Table~I.  \medskip

\begin{table}
\begin{tabular}{|c|c|c|c|}
\hline \hline
test point & particle   & mass (GeV)   & branching ratio  
\\
\hline \hline
S5         &  $\nn{1}$  &  122.          &   stable 
\\
\hline
S5         &  $\sbx{1}$  & 633.          & Br($\sbx{1} \to b \nn{1}$) = 26.6\%
\\
\hline
S5         &  $\sbx{2}$  & 663.          & Br($\sbx{2} \to b \nn{1}$) = 78.8\%
\\
\hline \hline
SPS1a      &  $\nn{1}$  &  97.          &   stable
\\
\hline
SPS1a      &  $\sbx{1}$  &  517.         & Br($\sbx{1} \to b \nn{1}$) = 4.4\%
\\
\hline
SPS1a      &  $\sbx{2}$  &  547.         & Br($\sbx{2} \to b \nn{1}$) = 29\%
\\
\hline \hline
L1         &  $\nn{1}$  &  97.8          &   stable
\\
\hline
L1         &  $\sbx{1}$  & 280.         & Br($\sbx{1} \to b \nn{1}$) = 40\%
\\
\hline
L1         &  $\sbx{2}$  & 354.          & Br($\sbx{2} \to b \nn{1}$) = 20\%
\\
\hline \hline
\end{tabular}
\label{t1}
\caption{Masses of sbottoms, the lightest neutralino, and branching ratios 
  of $\sbx{1,2} \to b\nn{1}$ for the test points considered in our numerical
  simulations.} 
\end{table}

In order to correctly treat all spin correlations we performed a parton--level
analysis including the UED interactions in MADGRAPH~\cite{madevent} and using
SMADGRAPH~\cite{smadgraph} for the SUSY simulation. In our calculations we
used CTEQ6L1 parton distribution functions \cite{Pumplin:2002vw} with the
factorization and renormalization scales are fixed by
$\mu_F=\mu_R=(\msb{1}+\msb{2})/2$ for the signal and
$\mu_F=\mu_R=m_V+\sum_i\p_{T_i}$ for the SM backgrounds, where $m_V$ is the
mass of the relevant electroweak gauge boson and $\sum_i\p_{T_i}$ is the sum
of the transverse momentum of additional jets.  \medskip

We simulated experimental resolutions by smearing the energies (but not
directions) of all final state partons with a Gaussian error.  We considered a
jet resolution $\Delta E/E = 0.5/\sqrt{E} \oplus 0.03$ and
$\sigma_{\sla{E}_T}=0.46\sqrt{\sum E_T}$ for the missing transverse energy
where $\sum E_T$ is the sum of the jet transverse energies. In addition to
that we included a b-tagging efficiency of $\varepsilon=60\%$ with a
mistagging probability of $1/200$ for light jets~\cite{btags}. No K-factors
were applied to signals or backgrounds since we do not expect QCD corrections
to change significantly the shape of kinematical
distributions~\cite{deq,prospino}.  Nevertheless, given the increasing
interest in studying spins correlations at the LHC by means of long and short
decay chains, it is important to check if this is indeed the case.

\section{ Analysis and  results}

\begin{figure}[tb]
  \begin{center}
\includegraphics[width=15.5cm]{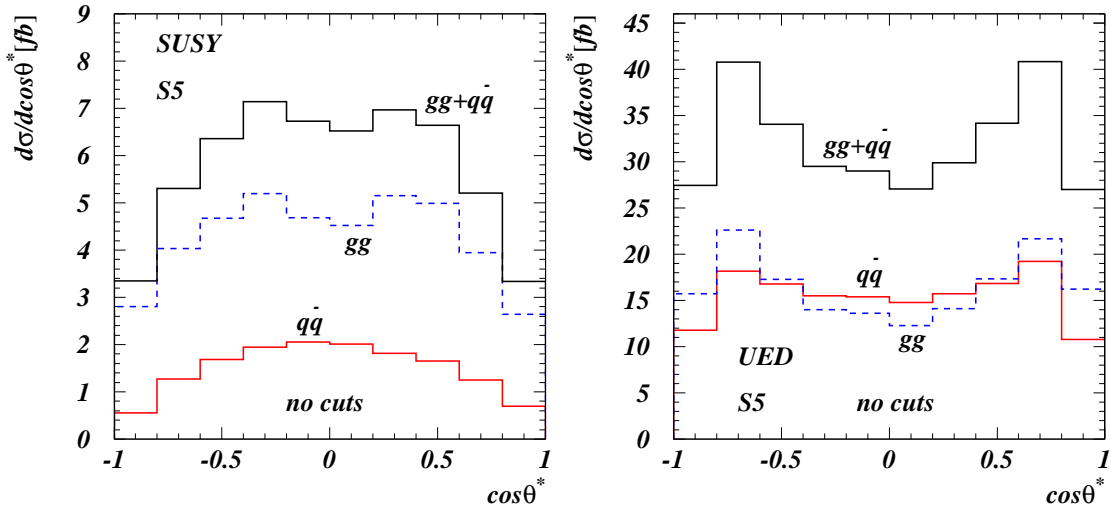}  
  \end{center}
  \vspace*{-8mm}
  \caption{Left panel: $\cos\theta^*$ distribution for the production of
    sbottoms coming from $q\bar{q}$ and $gg$ fusions. Right panel: same
    distribution as in the right panel but for the production of $b^{(1)}$.
    We used the test point S5 spectrum.}
\label{fig:qg}
\end{figure}

At the LHC we analyzed the production of sbottom pairs followed by their decay
in a $b$ and the LSP
\begin{equation}
  pp \to\sbx{1,2} \sbx{1,2}^* \to b \bar{b} \nn{1} \nn{1}
        \to b \bar{b} \sla{p}_T \; ,
\label{proc:lhc}
\end{equation}
\ie the signal is characterized by two $b$-tagged jets and missing transverse
energy.  One feature of the reaction (\ref{proc:lhc}) is that the s--channel
subprocesses $q \bar{q} \to \sbx{} \sbx{}^*$ present the well--known angular
distribution
\begin{equation}
   \frac{d \sigma}{d \cos\theta^*} \propto 1 - \cos^2\theta^*
\label{eq:polarsusy}
\end{equation}
where $\theta^*$ is the polar angle of produced scalar particles in their
center-of-mass frame.  However, this clean signature of the sbottom spin is
contaminated by the subprocess $ g g \to \sbx{} \sbx{}^*$ which contains
$t$-channel diagrams and quartic couplings; see Fig.~\ref{fig:qg} left panel.
Therefore, it is easier to decipher the new state spin if we enhance the
importance of the $q\bar{q}$ s--channel subprocesses via a judicious choice of
cuts.  Notwithstanding, the cuts to isolate the signal must not introduce bias
in the angular distributions used to study the sbottom spin.

\begin{figure}[tb]
  \begin{center}
    \includegraphics[width=7.3cm]{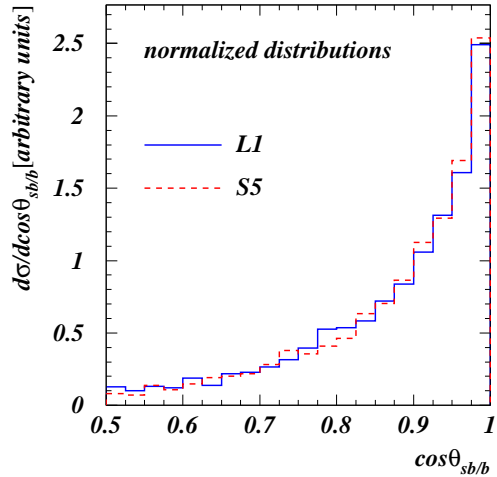}
  \end{center}
  \vspace*{-8mm}
  \caption{ Cosine of the opening angle between the bottom and the sbottom in
    center--of--mass system for two different mass spectra considered
    here.}
\label{fig:t1t2}
\end{figure}

On the other hand, the center--of--mass angular distribution of KK bottoms in
UED produced by $q \bar{q}$ fusion is
\begin{equation}
   \frac{d \sigma}{d \cos\theta^*} \propto 1
   +\left(\frac{E^2_{b_1}-m^2_{b_1}}{E^2_{b_1}+m^2_{b_1}}\right)
   \cos^2\theta^* \; ,
\label{eq:polarued}
\end{equation}
where $M_{b_1}$ and $E_{b_1}$ are the mass and energy respectively of the
$b^{(1)}$ in the center--of--mass frame. This distribution peaks in the
forward and backward directions being quite distinct of the SUSY prediction.
Moreover, we must also include the $gg\rightarrow b^{(1)}\bar{b}^{(1)}$ which
contains t- and s-channel contributions which present a peak towards the
forward and backward regions as well; see Fig.~\ref{fig:t1t2} right panel.
\medskip


%

At the LHC we can not reconstruct the the polar angle ($\theta^*$) of produced
particles in their center-of-mass frame due to the presence of undetected
$\nn{1}$ or $\gamma^{(1)}$. Therefore, we must use an alternative variable
that retains part of the information carried by $\theta^*$.  A convenient
variable to use in our analysis is~\cite{barr2}
\begin{equation}
   \cos\theta_{bb}^*\equiv \tanh\left ( \frac{\Delta\eta_{bb}}{2} \right)
\label{eq:barr}
\end{equation}
where $\Delta\eta_{bb}$ is the rapidity separation of the $b$-tagged jets.
Notice that $\Delta\eta_{bb}$ is invariant under boosts along the collision
axis, and consequently, $\cos\theta_{bb}^*$ is invariant under boosts as well.
The angle $\theta_{bb}^*$ is the polar angle between each reconstructed bottom
jet direction in the longitudinally boosted frame in which the rapidities of
the bottoms are equal and opposite.
\medskip

In order to $\cos\theta_{bb}^*$ carry information of the produced
particle spin, the flight directions of sbottoms and bottoms must be
correlated.  We depict in the left panel of Fig.~\ref{fig:t1t2} the cosine of
the opening angle between the bottom and the sbottom flight directions in the
$\sbx{} \sbx{}^*$ center--of--mass system for the S5 and L1 spectra. Clearly,
the bulk of the signal is concentrated in region of small opening angles as a
consequence of the large energy of the sbottoms after cuts.  \medskip

\begin{figure}[t]
\begin{center}
\includegraphics[width=15.5cm,angle=0]{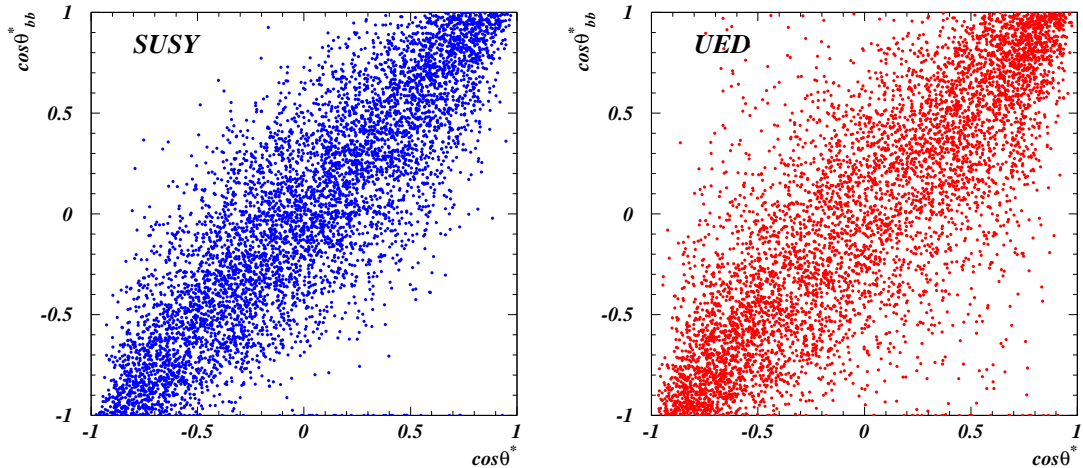} 
\end{center}
\vspace*{-8mm}
\caption{In the left (right) panel we display a scattered plot
    $\cos\theta_{bb}^* \otimes \cos\theta^*$ for sbottoms ($b^{(1)}$)
    production. Here we chose the parameter point S5.}
\label{figlhc:corrs5}
\end{figure}

Fig.~\ref{figlhc:corrs5} shows that $\cos\theta_{bb}^*$ is indeed strongly
correlated to the cosine of production polar angle ($\theta^*$) of the
sbottoms and $b^{(1)}$'s in their center--of--mass system.  Therefore, we must
expect the shape of $\cos\theta_{bb}^*$ distributions to resemble the polar
angle spectra of sbottoms and KK bottoms apart from some smearing effects due
to non-perfect correlations between the bottoms and sbottoms (KK bottoms)
flight directions.  Nevertheless, a clear separation between SUSY and UED
distributions should be possible as was demonstrated in the case of smuon pair
production at the LHC~\cite{barr2}.  Taking a closer look at
Fig.~\ref{figlhc:corrs5} we already realize that UED events are slightly more
concentrated near $\cos\theta_{bb}^*=\pm 1$ while SUSY events are
homogeneously distributed along the direction $\cos\theta_{bb}^* = \cos
\theta_{\tilde{b}}$.  \medskip

In our analysis, we included the following backgrounds for the process
$ pp \to b \bar{b} \sla{p}_T$:

\begin{itemize}

\item SM QCD and electroweak production of $b \bar{b} Z$ with $Z\to \nu
  \bar{\nu}$ that accounts for $\simeq 91$\% of the total background
    after cuts for the S5 and SPS1a test points and $\simeq 72$\% for the L1
  scenario.

\item Reducible electroweak and QCD backgrounds like $jjZ$, $jjW$,
  $j\tau\nu_\tau$, $b \bar{b} j$, $jj$, $jjj$, and $jjb$ where some of the
  decay products evade detection and $j$ stands for a light jet that might be
  mistagged as a $b$ jet. These dangerous backgrounds are efficiently reduced
  by the requirement of two $b$-tagged jets.  The second largest background
   after cuts is $bbW$ where the lepton from the W boson has a large
  rapidity $|\eta_\ell| > 2.5$.


\item SUSY processes, excluding $\sbx{} \sbx{}^*$ production, that lead to the
  final state $b \bar{b} \nn{1} \nn{1}$. 
\end{itemize}

In order to properly trigger~\cite{triggers} the event and tag the
b-jets~\cite{btags} the following acceptance cuts were imposed in all cases
\begin{equation}
\begin{array}{llll}
  |\eta_b| < 2.5\;\;\;\;,\;\;\;\; & p_{T}^{b} >  100
   \gev\;\;\;\;,\;\;\;\; & \Delta R_{bb} >  0.7\;\;\;\;,\;\;\;\; 
  &  \sla{p}_T  >  100\;\;\;\;. 
\end{array}
\label{lhccuts:acept}
\end{equation}

\bigskip

A potentially large background is the QCD production of dijets once we take
into account that mismeasurements of the jets properties can lead to missing
transverse momentum and that this process has a huge cross section. This
background can be efficiently reduced by the missing transverse momentum cut
Eq.~(\ref{lhccuts:acept}) and by requiring that the azimuthal angle between
the jets and the missing transverse momentum satisfy
\begin{equation}
|\Delta\Phi(\sla{p}_T,p_{Tj})| >  0.3\;\;\;\;. 
\label{lhccuts:dijets}
\end{equation}

In the L1 test point simulations we applied the following cuts not only to
enhance the signal and deplete the background but also to augment the
importance of the $q \bar{q}$ s-channel subprocess
\begin{equation}
\begin{array}{lll}
   M_{b\bar{b}}  >  300 \gev\;\;\;\;,\;\;\;\;  & M_{\rm eff}  >  
   600 \gev\;\;\;\;,\;\;\;\; & |\Delta\Phi(\sla{p}_T,p_{Tj_{soft}})| <
   2.4\;\;\;\;,  
\end{array}
\label{lhccuts:L1}
\end{equation}
where $M_{\rm eff}$ is the sum of all jet and missing transverse momenta and
$\Delta\Phi(\sla{p}_T,p_{Tj_{soft}})$ the azimuthal angle between the missing
transverse momentum and the softest jet. After cuts the SUSY signal cross
section is 38.1 fb while the background cross section is 2.4 fb, leading to
$S/B \simeq 16.6$.
\medskip

For the parameter point S5 we used the following cuts to optimize the signal
\begin{equation}
\begin{array}{lll}
   M_{b\bar{b}}  >  600 \gev\;\;\;\;,\;\;\;\;  & M_{\rm eff}  >   1 \tev
   \;\;\;\;,\;\;\;\; & |\Delta\Phi(\sla{p}_T,p_{Tj_{soft}})| <  2.4\;\;\;\;. 
\label{lhccuts:s5}
\end{array}
\end{equation}
The SUSY signal cross section at this test point is 4.55 fb with a background
of 1.24 fb and $S/B \simeq 3.7$ while the $q\bar{q}$ fusion accounts for
$\simeq 40$\% of the signal events. On the other hand, $\simeq 45$\% of the
UED signal stems from $q \bar{q}$ fusion.  The cuts employed in this case were
harder for two reasons: first, the bottoms are harder in this case once the
sbottom is much heavier than the neutralino compared to L1 case; second, the
signal cross section is significantly smaller than in the L1 case which
required a deeper suppression of backgrounds to avoid a severe bias on
  the signal angular distributions.  \medskip

Finally, we imposed the following cuts for the test point SPS1a
\begin{equation}
   M_{b\bar{b}}  >  600 \gev\;\;\;\;,\;\;\;\;   M_{\rm eff}  >  1 \tev\;\;\;\;, 
\label{lhccuts:sps1}
\end{equation}
which lead to a signal (background) cross section of 1.07 (1.46) fb and $S/B
\simeq 0.73$.  \medskip

\begin{figure}[tb]
  \begin{center}
  \includegraphics[width=8.5cm]{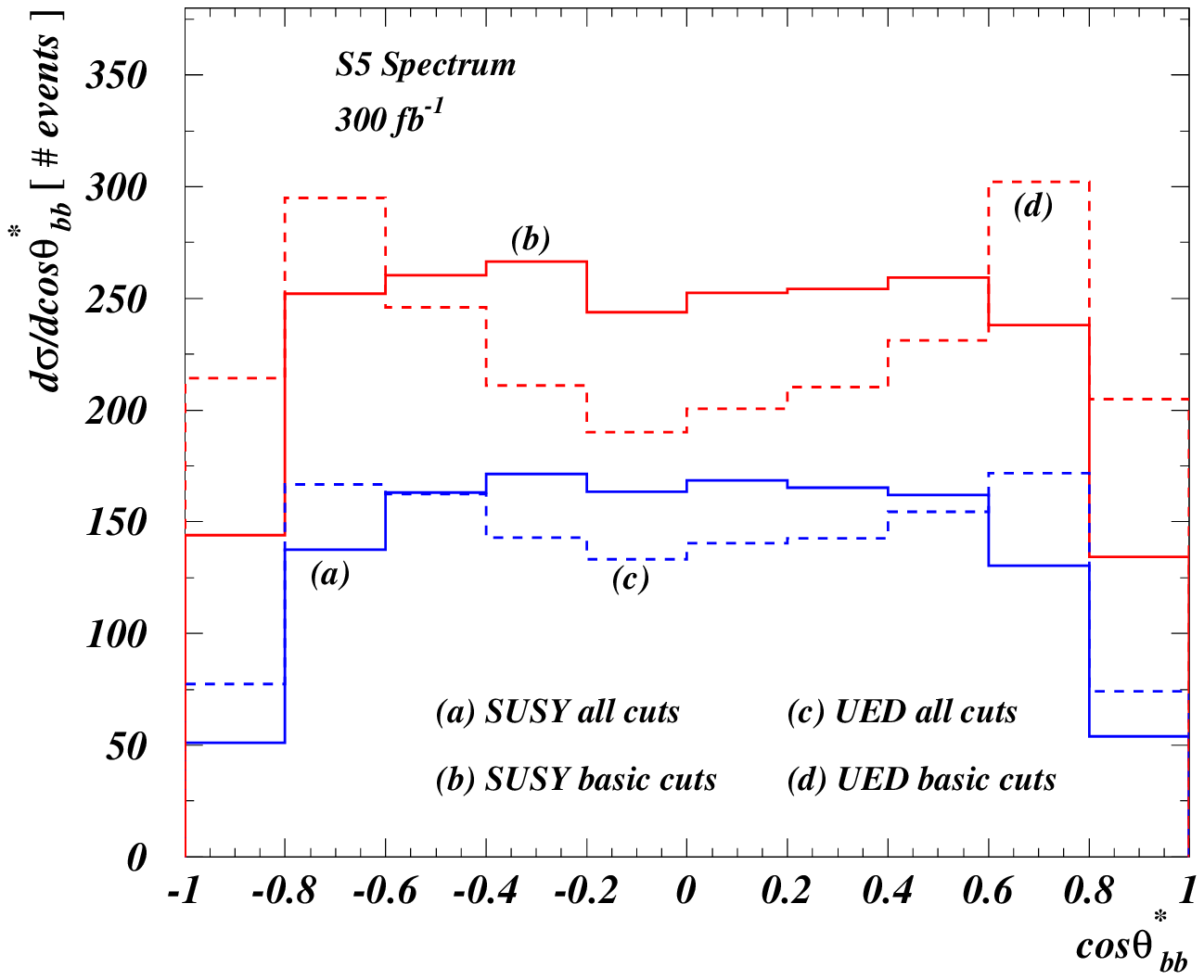}
  \includegraphics[width=8.5cm]{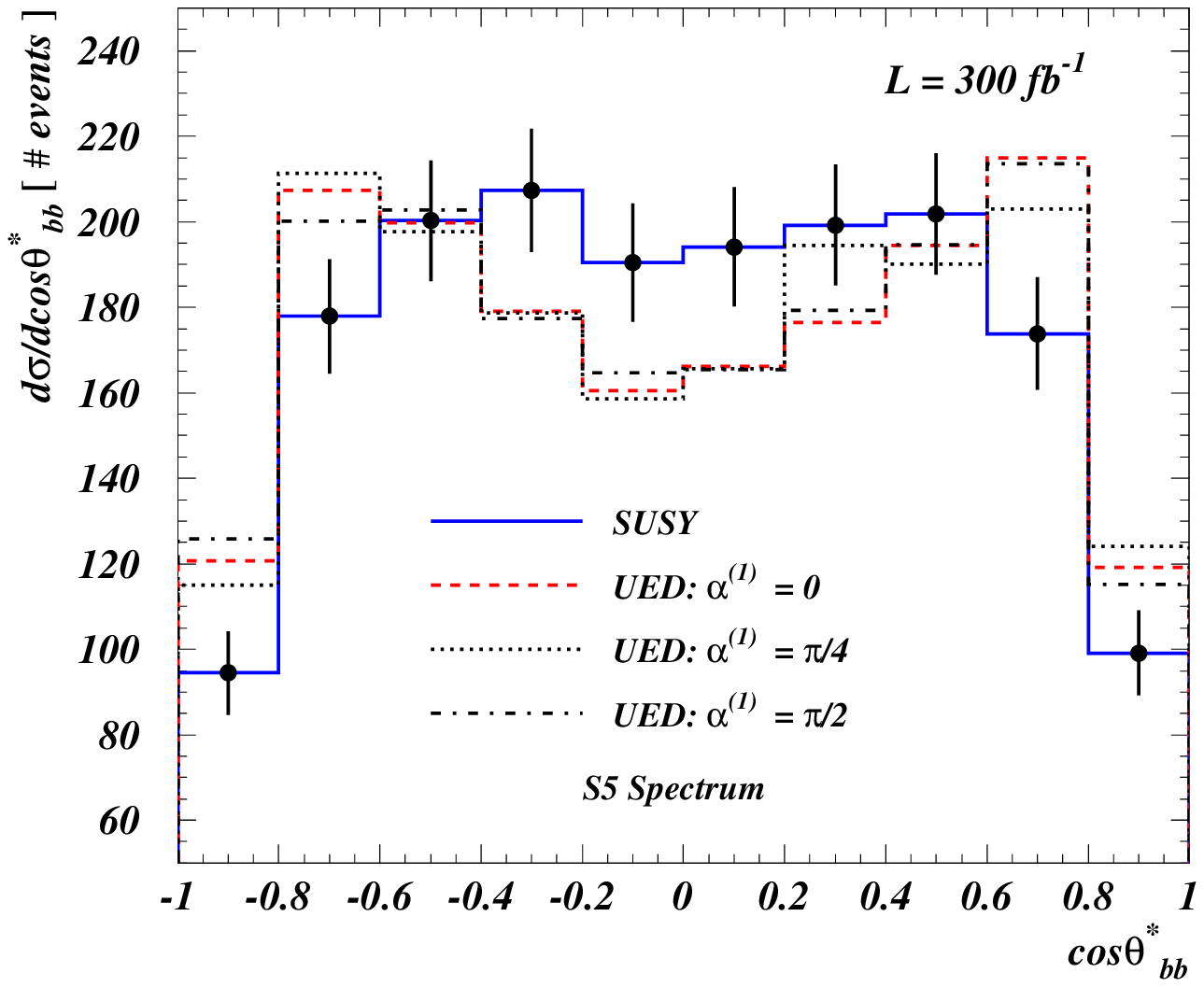}
  \end{center}
  \vspace*{-8mm} \caption{Left panel: impact of cuts on the
$d\sigma/d\cos\theta^*_{bb}$ distributions for SUSY and UED signals. We
normalized the UED signal cross sections to the SUSY ones using the S5
spectrum.  Right panel: SUSY and UED $d\sigma/d\cos\theta^*_{bb}$
distributions without backgrounds but varying the KK bottom mixing angle
$\alpha^{(1)}$. The basic cuts are the acceptance defined in
Eq.~(\ref{lhccuts:acept}) plus $M_{b\bar{b}}  >  600 \gev$.}
\label{figlhc:s5mix} 
\end{figure}

In the left panel of Fig.~\ref{figlhc:s5mix} we show the impact of cuts on
the $d\sigma/d\cos\theta^*_{bb}$ distributions for SUSY and UED predictions
assuming the S5 spectrum. We normalized the UED signal cross section to the
SUSY one. As the background events populate the bins of larger
$|\cos\theta^*_{bb}|$, the selection cuts tend to suppress that region
enhancing the central region for both SUSY and UED. The variable
$\Delta\Phi(\sla{p}_T,p_{Tj})$ is efficient in rejecting dangerous SM
backgrounds like dijet production, however, it has a potential to bias the
distributions as we can see in the left panel of Fig.~\ref{figlhc:s5mix}.
Therefore, a harder cut on this variable is not recommended in the study of
spin correlations. \medskip

A natural question at this point is whether we can mimic the SUSY signal by
varying the UED mixing angles. According to Section~\ref{sec:ued} there is
not very much room to modify the UED Lagrangian to bring kinematical
correlations closer to the SUSY prediction.  The KK weak mixing angle
$\theta_w^{(n)}$ in Eq.~(\ref{eq:weinberg}) is fixed by the KK masses, so we
can not change it while keeping the masses fixed.  The same limitations hold
when we try to adjust the mixing between the singlet and doublet KK
fermions, described by the angle $\alpha^{(n)}$; see Eq.~(\ref{eq:eigen}).
In contrast to the 3rd--generation sfermion sector in the MSSM, the UED
mixing angle is not a (third) free parameter, even if we move around the
masses invoking boundary conditions.  Notwithstanding, for illustration
purpose we vary $\alpha^{(1)}$ in the right panel Fig.~\ref{figlhc:s5mix} to
check whether the SUSY $\cos\theta_{bb}^*$ can be reproduced by a UED decay
chain with different couplings to the fermions. From
Eq.~(\ref{eq:lagrangian1}) we see that varying $\alpha^{(1)}$ effectively
enhances the left or right couplings of the KK bottom decay into bottom plus
LKP.  This figure allows us to see that the changes in the UED
parameters are not sufficient to mimic the SUSY predictions.  \medskip

We depict in Fig.~\ref{figlhc:s5} the $\cos\theta_{bb}^*$ spectrum
with/without adding the background for the S5 test point, an integrated
luminosity of 300 fb$^{-1}$, and after applying cuts (\ref{lhccuts:acept})
and (\ref{lhccuts:s5}). To avoid using any information
but the spin we assume the S5 spectrum for the UED particles and normalize
their production cross section times branching fractions to the SUSY rate.
From Figure~\ref{figlhc:s5}, we can easily see that the production of
fermionic strongly interacting states (UED) favor large separations
($\cos\theta_{bb}^*$) between the $b$-tagged jets while the production of
scalar particles (SUSY) leads to a rather constant distribution. Note that
these distributions indeed resembles the distributions of the production
angles in the center--of--mass system which reflects the correlation between
these observables. Moreover, the UED $\cos\theta_{bb}^*$ distribution is
similar to the SM background one since they take place through similar
diagrams containing KK partners or SM particles with the same spin quantum
numbers. 
\medskip

For the assumed integrated luminosity it is rather easy to distinguish between
the two models.  These two possibilities can be disentangled, for instance,
through the asymmetry:
\begin{equation}
A=\frac{\sigma(|\cos\theta^*_{bb}|<0.5) - \sigma(|\cos\theta^*_{bb}|>0.5)}
 {\sigma(|\cos\theta^*_{bb}|<0.5)+\sigma(|\cos\theta^*_{bb}|>0.5)} \; .
\label{asym}
\end{equation}
This asymmetry is $+0.238 \pm 0.023$ for the UED spin assignment while the
SUSY interpretation it is significantly larger $+0.373 \pm 0.022$ where the
quoted errors are statistical. We estimate that an integrated luminosity of
$300\ifb$ is needed to reach a 5$\sigma$ level signal for the S5
spectrum.  \medskip

\begin{figure}[tb]
  \begin{center}
  \includegraphics[width=8.5cm]{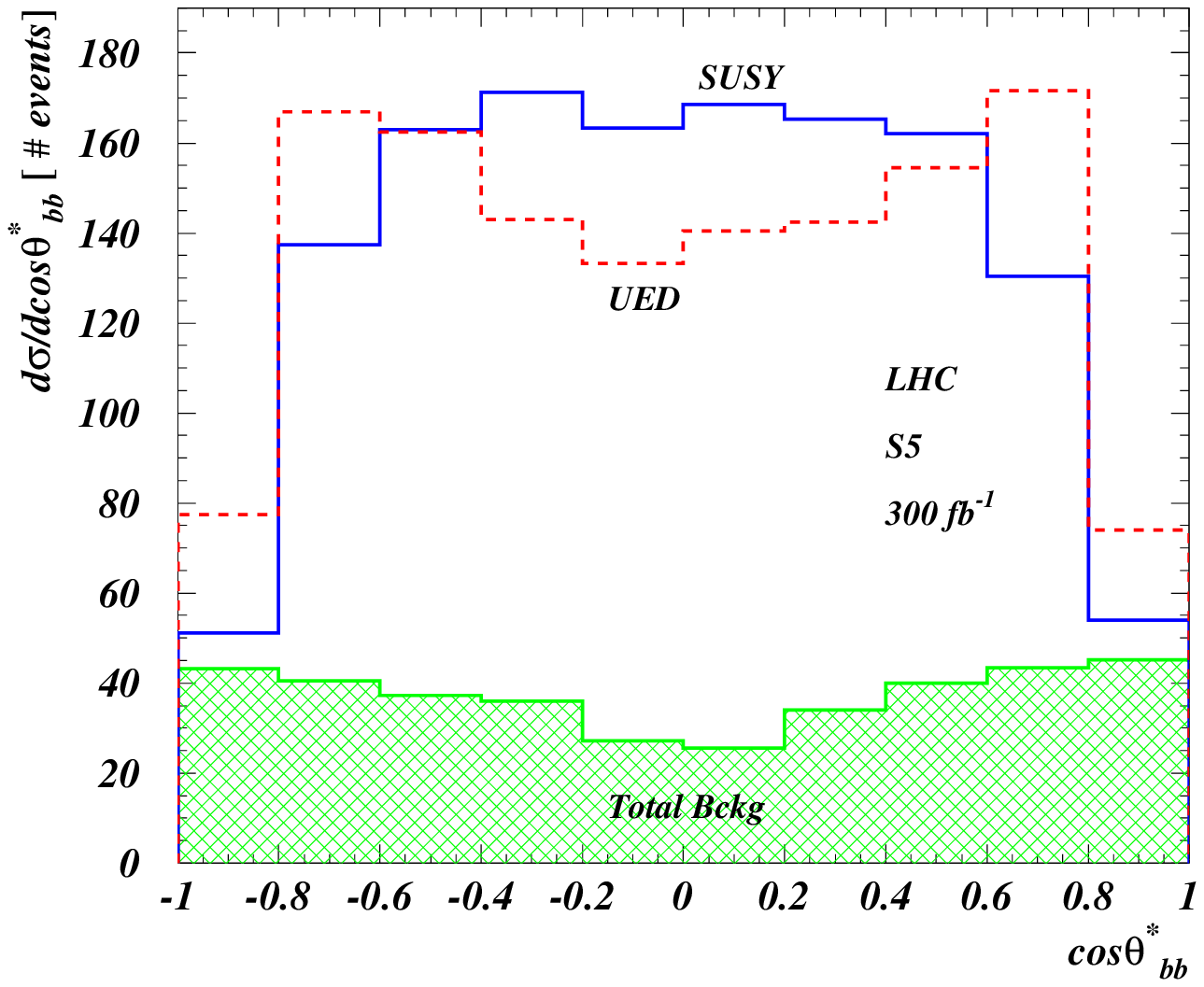} 
  \includegraphics[width=8.5cm]{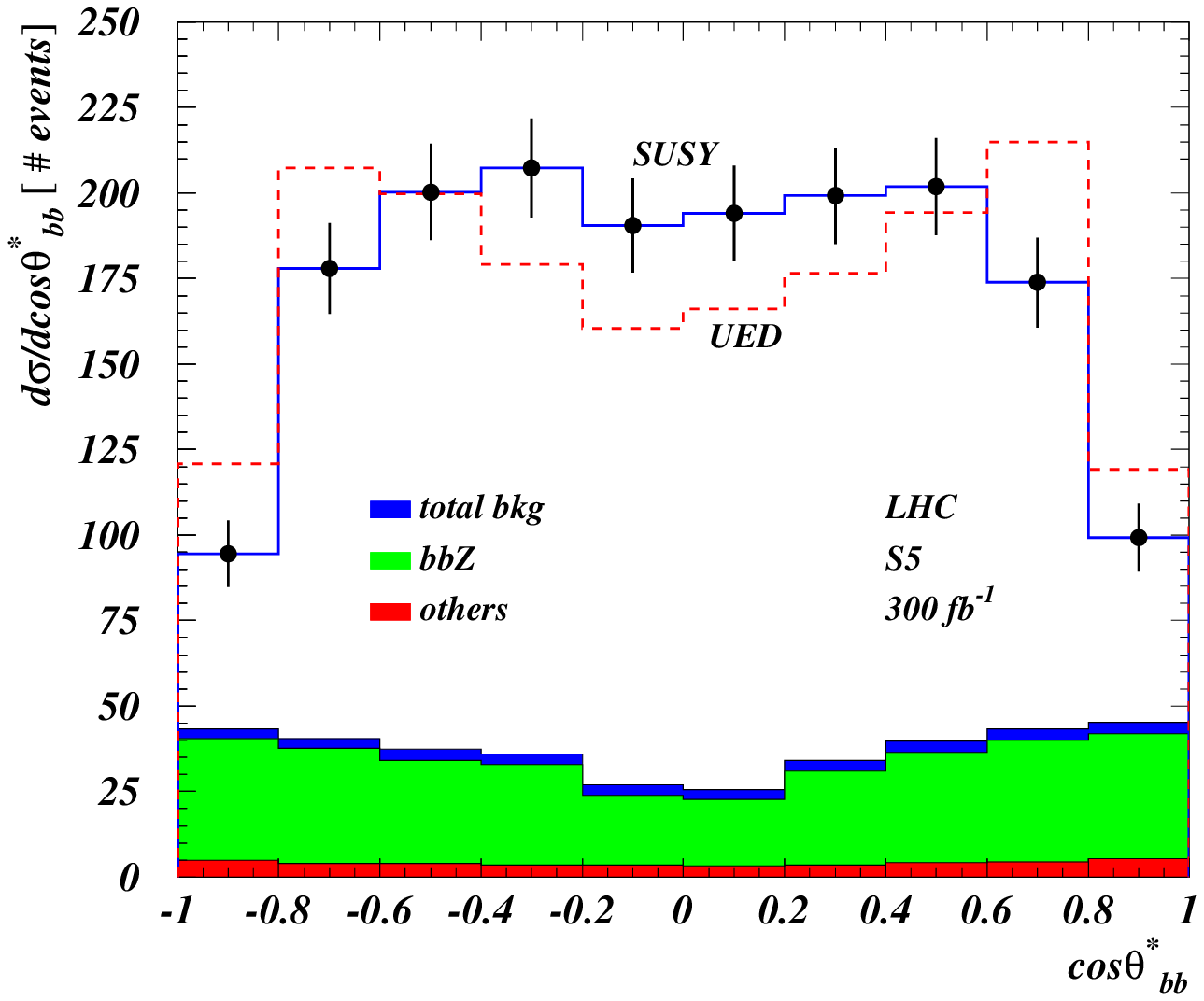}
  \end{center}
  \vspace*{-8mm}
  \caption{In the left (right) panel we plot the $\cos\theta_{bb}^*$
    distribution without (with) the addition of the background contribution.
    Here we used the parameter point S5 and assumed an integrated luminosity
    of 300 fb$^{-1}$. The error bars indicate the expected statistical
    uncertainties.}
\label{figlhc:s5}
\end{figure}
\begin{figure}[tb]
  \begin{center}
  \includegraphics[width=8.5cm]{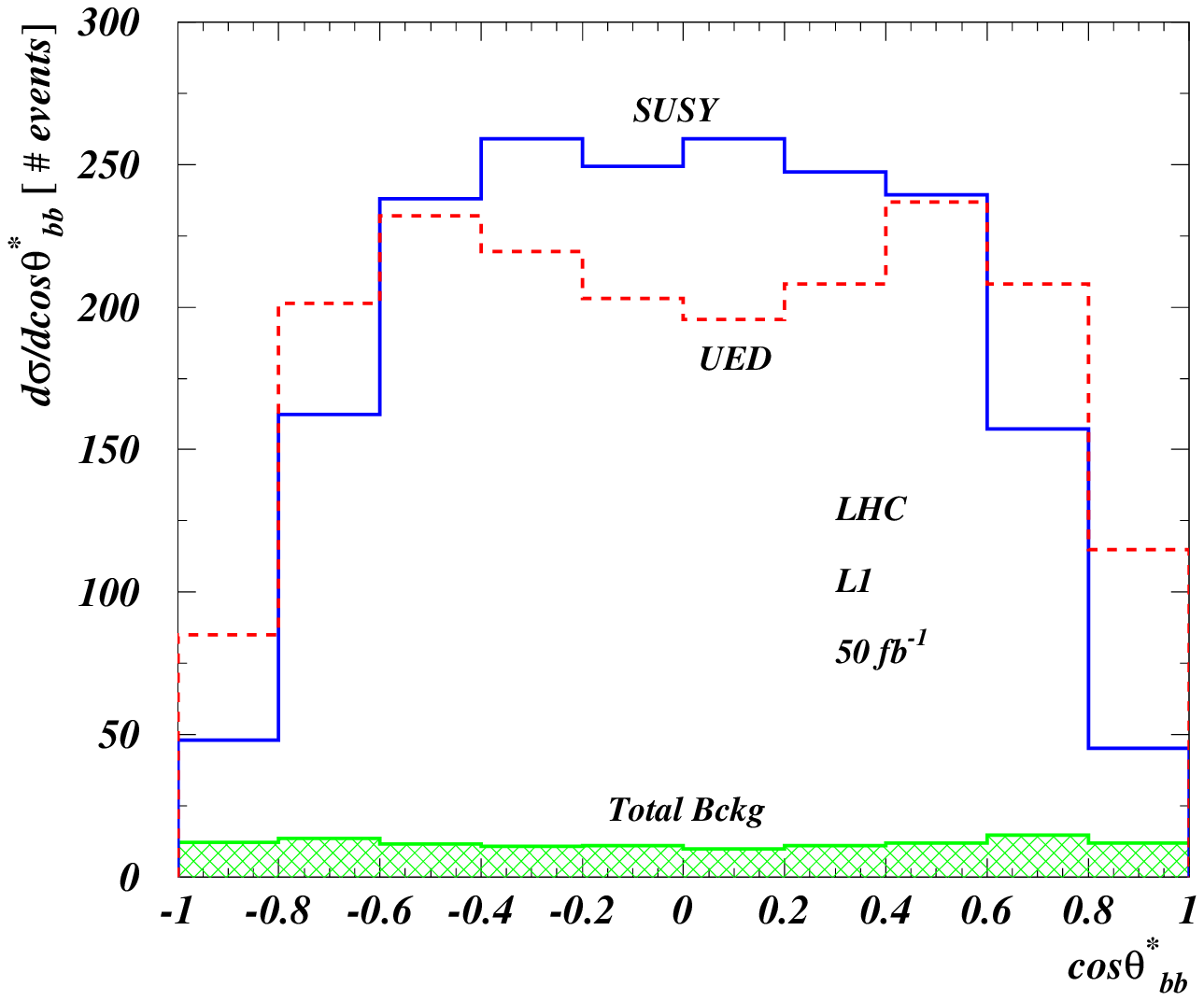} 
  \includegraphics[width=8.5cm]{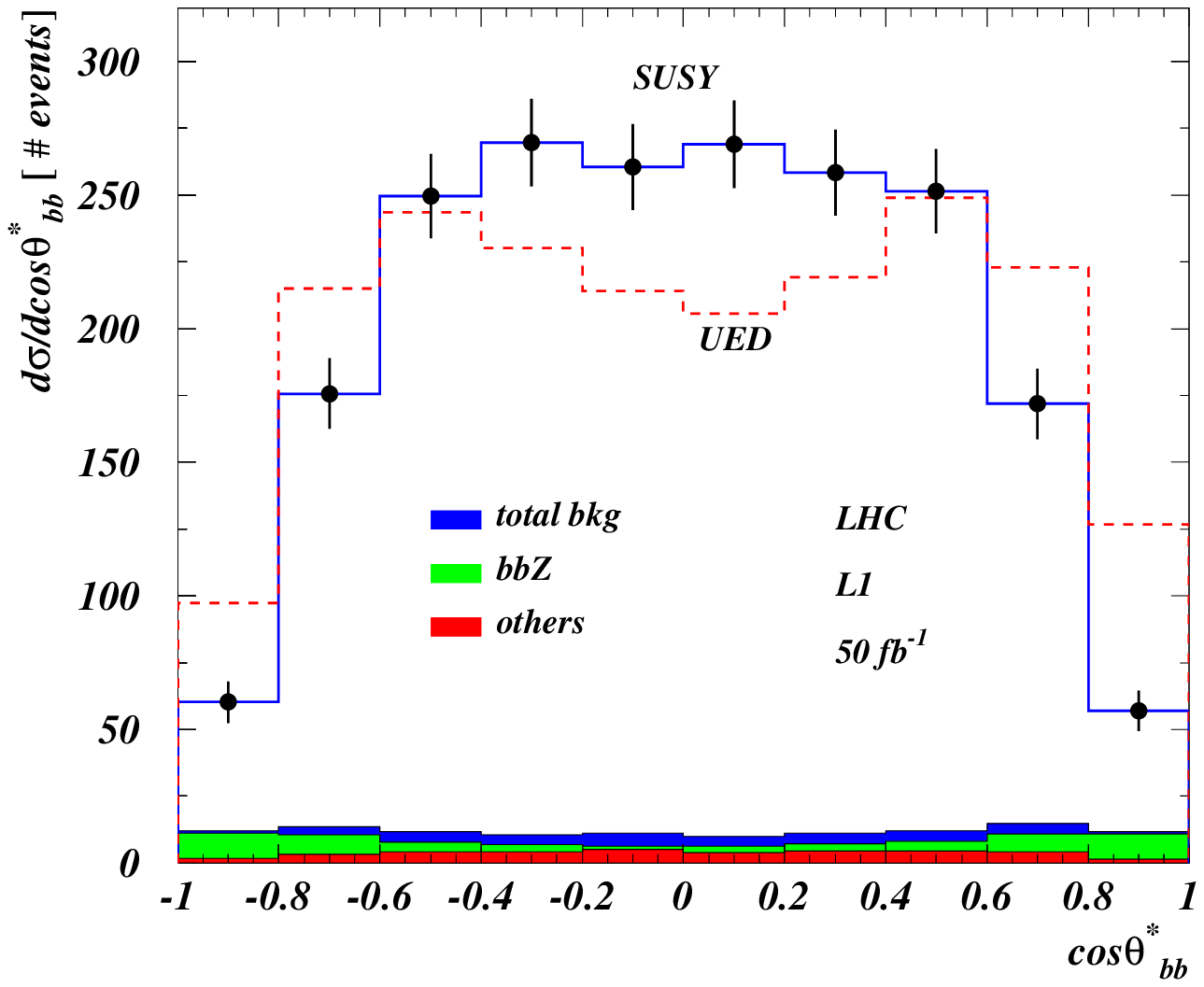}
  \end{center}
  \vspace*{-8mm}
  \caption{Same as Fig.~\ref{figlhc:s5} but for the parameter point L1 and an
    integrated luminosity of $50 \ifb$.}
\label{figlhc:l1}
\end{figure}

The determination of the sbottom spin is much easier for the test point L1 due
to the large $\sbx{} \to b \nn{1}$ branching ratio and production cross
section. We depicted in Fig.~\ref{figlhc:l1} the $\cos\theta^*_{bb}$
distribution and without/with adding the SM background. In this case a mere
integrated luminosity of $15 \ifb$ is enough to
discriminate at the 5$\sigma$ level UED and SUSY. The asymmetries are given by
$+0.565\pm 0.019$ for SUSY and $+0.365\pm 0.021$ for UED  for this
integrated luminosity.  Therefore, there is a clear distinction
between the two spin assignments  (UED $\times$ SUSY) for both S5 and
L1 mass spectra and the above integrated luminosities.  \medskip

\begin{figure}[tb]
  \begin{center}
  \includegraphics[width=8.5cm]{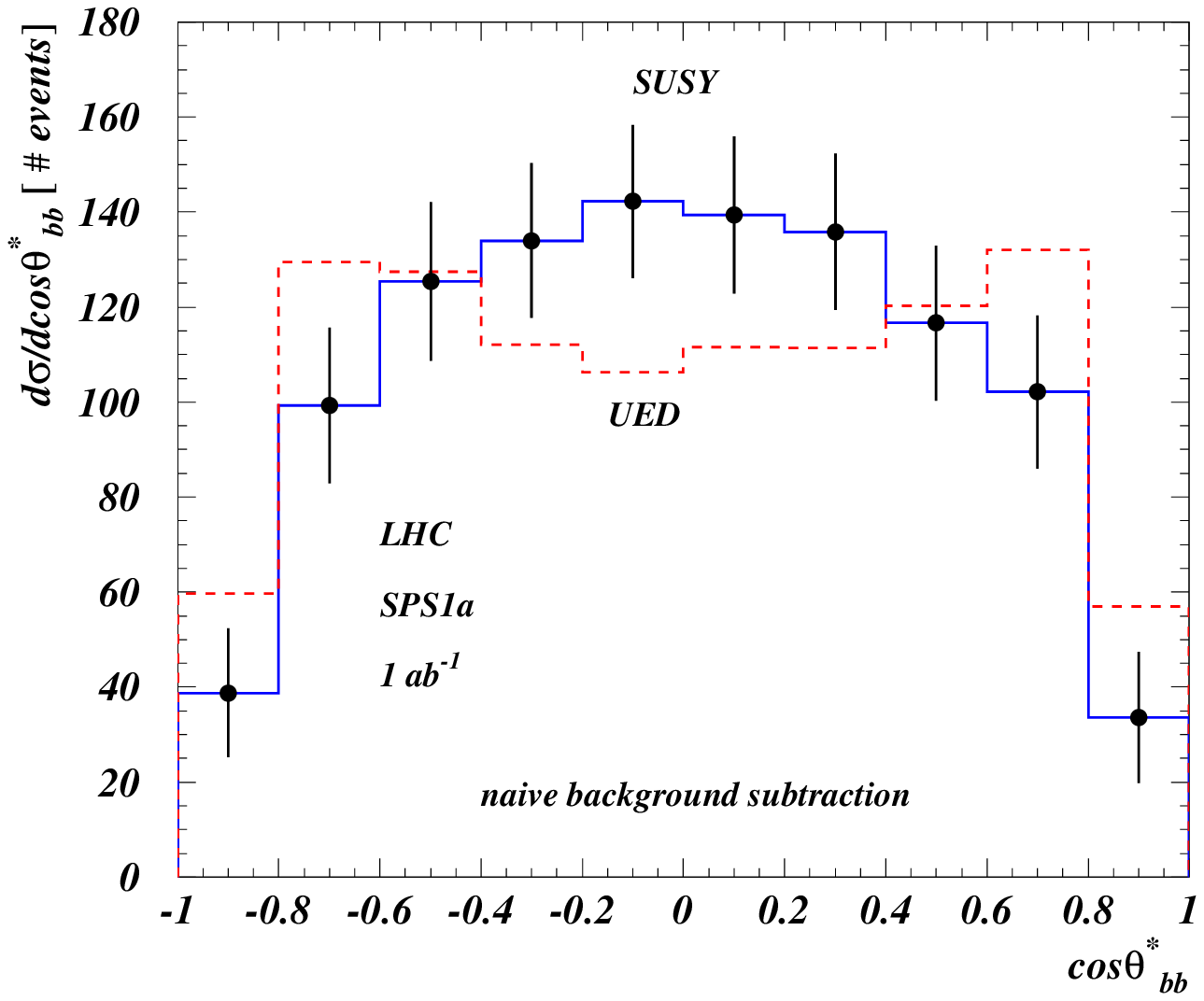} 
  \includegraphics[width=8.5cm]{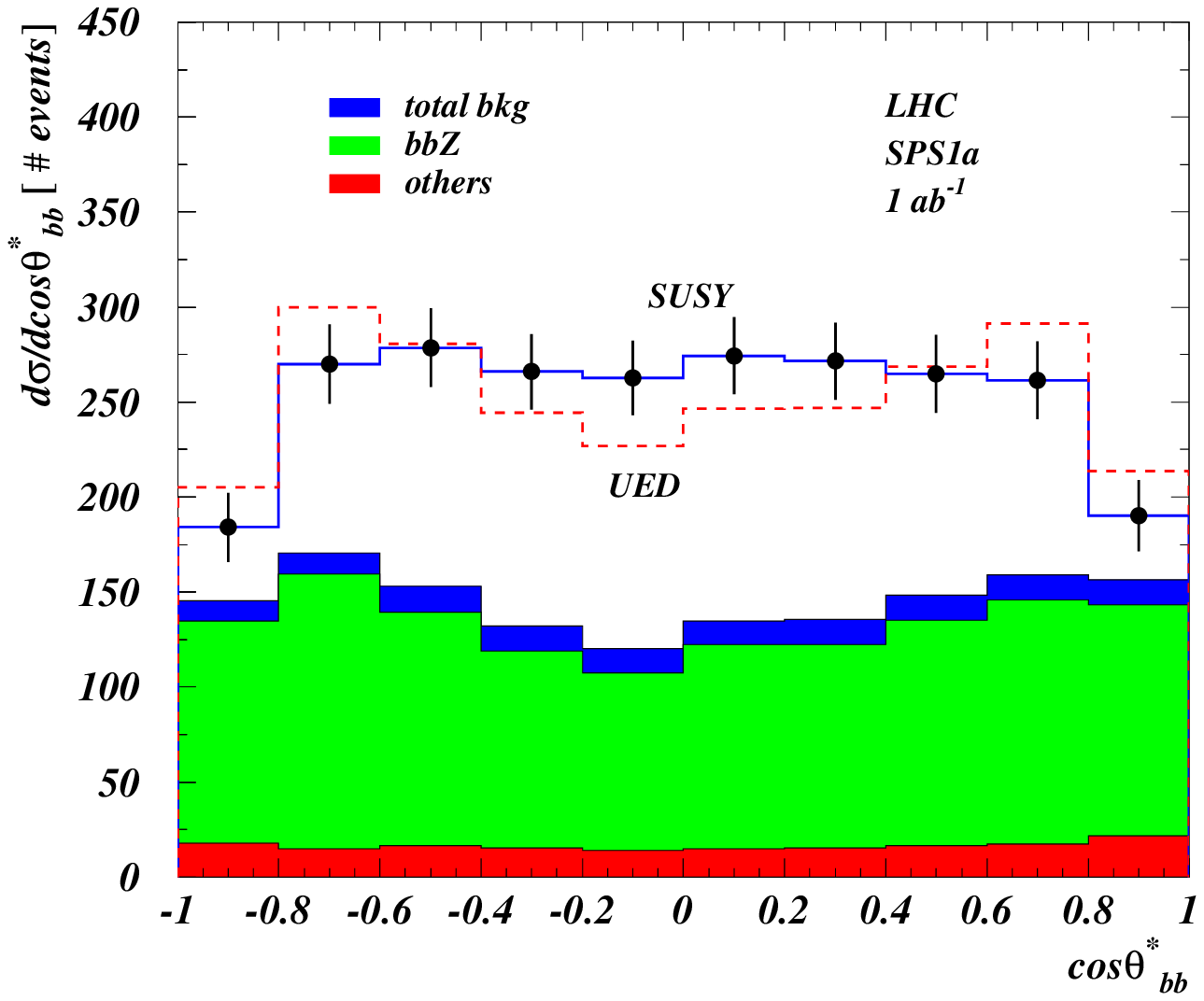}
  \end{center}
  \vspace*{-8mm}
  \caption{Same as Fig.~\ref{figlhc:s5} but for the parameter point SPS1a and
    an integrated luminosity of $1 \hbox{ ab}^{-1}$.}
\label{figlhc:sps1a}
\end{figure}

The results for the reference point SPS1a are quite different from the S5 and
L1 ones. Due to small $\sbx{} \to b \nn{1}$ branching ratio, the SM
backgrounds play a major role. We can see in the left panel of
Fig.~\ref{figlhc:sps1a} that the pure SUSY and UED $\cos\theta^*_{bb}$
spectrum are quite different.  However, once we add the SM background, which
has a shape similar to UED, it is no longer easy to separate the SUSY from
UED, even for an integrated luminosity of 1 ab$^{-1}$; see the right panel of
Fig.~\ref{figlhc:sps1a}. For instance, the asymmetry (\ref{asym}) is $0.282
\pm 0.019$ for SUSY and $0.200 \pm 0.019$ for UED and this
extremely large luminosity. In this case SUSY and UED can be discriminated
only at at $\sim 4\sigma$ level.  Therefore, it is important for this point to
subtract the SM background which can be estimated, for instance, from the
measurement of the $b \bar{b} \mu^+ \mu^-$ cross section.  If we neglect
systematic errors associated to this extraction from data, the sbottom spin
can be determined at 5$\sigma$ level for an integrated luminosity of $\simeq
500 \ifb$. The left panel of Fig.~\ref{figlhc:sps1a} displays the
distributions of SUSY and UED after a naive background subtraction where we
just did not add the background to the signals and the error bars calculated
as $\sqrt{S+B}$. A dedicated study is necessary to determine the actual impact
of the statistical and systematic errors~\cite{subtraction}. \medskip

In general, we noticed that the expected shape of distributions in
$\cos\theta^*_{bb}$ are similar to the ones in smuon pair production and
decay to $\mu\chi^0_1$~\cite{barr2}, which shows the universality and
robustness of the method. Moreover, in both cases ($\sbx{}^* sbx{}$ and
$\smuon{}^*\smuon{}$) the S5 spectrum seems to be more promising as compared
to the SPS1a spectrum, for example, SUSY {\it versus} UED discrimination of
smuon spin assignments is possible in the case of S5 (SPS1a) for an
integrated luminosity of $200 \ifb$($500 \ifb$).

As a final remark note that, apart from negligible effects from $b\bar{b}$
initiated contributions which include interactions with gluinos and
electroweak interactions with $Z$ bosons, the production rate does not depend
upon the left or right nature of the sbottom since the QCD interactions to the
gluons are blind to these details. On the other hand, the
sbottom--bottom--neutralino vertex is sensitive to the relative content of
mass eigenstates in terms of left and right states which by its turn depend
upon the mass spectrum. As we have pointed out in the Section~\ref{sec:test}
the lightest sbottom is almost entirely a left state in the S5 mass spectrum,
while it is an equal mixture of left and right states in the L1 mass
spectrum.  As can be seen in the Figs.~\ref{figlhc:s5} and~\ref{figlhc:l1}, it
seems plausible to conclude that the distributions are not sensitive to the
particular mixtures of left and right states of the sbottom mass eigenstates,
however a more detailed study including more test points is necessary to
confirm this indication.

\section{Conclusions}

In the near future the LHC will start its endeavor in searching for new
physics signals. Once those signals have been identified as the production of
new states the next logical step will be the determination of the underlying
model among all candidates by measuring the properties and interactions of the
new particles. The size of production cross sections and mass spectra might
provide valuable hints about the underlying model, nevertheless, an undisputed
discrimination will only be possible after determination of the spins of the
new states. We showed that the discrimination between a SUSY spin
interpretation against an UED one is possible in the case of scalar bottom
(fermionic KK bottom) pair production in the reaction $pp \rightarrow b
\bar{b}\sla{p}_T$.  Using the variable proposed in
Ref.~\cite{barr2}, see Eq.~(\ref{eq:barr}), we demonstrated that a clear
determination of the spin of the decaying strongly interacting particle is
possible provided the production cross sections and branching ratios into
$b\nn{1}$ are sufficiently large.Even in the worst scenario studied, the
SPS1a spectrum, the determination of spins might be possible after background
subtraction, however, a dedicated study of the impact of statistical and
systematic errors on the spin determinations will be necessary in this
  test point.

\begin{acknowledgments}
  We would like to thank Tilman Plehn for insightful comments and
  Fabio Maltoni and the CP3 team for their help with the implementation
  of UED model into Madgraph. This research was supported in part by
  Funda\c{c}\~{a}o de Amparo \`a Pesquisa do Estado de S\~ao Paulo
  (FAPESP) and by Conselho Nacional de Desenvolvimento Cient\'{\i}fico
  e Tecnol\'ogico (CNPq).
\end{acknowledgments}



\end{document}

\bibitem{Meade:2006dw}
P.~Meade and M.~Reece,
arXiv:hep-ph/0601124.
